\documentclass{aa}
\usepackage{graphicx}

\begin{document}

 \title{RXTE broadband X-ray spectra of intermediate polars and white dwarf
mass estimates    }

 \author{
    V. Suleimanov\inst{1,2}
       \and
    M. Revnivtsev\inst{2,3}
       \and
    H. Ritter \inst{2}
          }

 \offprints{V. Suleimanov:  vals@ksu.ru}

 \institute{
            Kazan State University, Kremlevskaya
            str.18, 420008 Kazan, Russia
         \and
            Max-Planck-Institut f\"ur
            Astrophysik, Karl-Schwarzschild-Str. 1,
            D-85740 Garching bei M\"unchen, Germany
         \and
            Space Research Institute, Russian Academy of
            Sciences, Profsoyuznaya 84/32, 117810 Moscow, Russia
              }

\date{Received May 13, 2004}

\titlerunning{RXTE broadband
X-ray spectra of IPs and white dwarf mass estimates}

\abstract{We present
results of an analysis of broadband spectra of 14  intermediate polars
obtained with the RXTE observatory (PCA  and HEXTE spectrometers, 3-100
keV). By means of our calculations of the  structure and the emergent
    spectrum of the post-shock region of intermediate polars we fitted
    the observed spectra and obtained estimates of the white dwarf masses. The
estimated masses are compared with masses obtained by other authors.
}

\maketitle

\section{Introduction}

Intermediate polars (IP) form a
sub-class of magnetic cataclysmic variables (mCV) (Warner \cite{Wr}).
A white dwarf (WD) accretes  matter from a companion star (typically a
red dwarf) that fills its Roche  lobe. This matter forms an accretion
disc inside the Roche lobe of the primary which, however, is
disrupted by the WD magnetic field at some distance from the WD
surface. As a result the accreting matter freely falls on to the WD
surface and  forms a strong shock near its surface. The post-shock
matter has a high temperature ($\sim$ 10--20 keV) and emits X-rays via
optical thin thermal plasma emission (Lamb \& Masters \cite{LM}).

The temperature of the post-shock matter depends
on the WD mass and therefore the X-ray spectra  of IPs can be used for WD
mass determination (Rothschild et al. \cite{Rot}).  Ishida  (\cite{Ish})
 had estimated the masses of the WDs of IPs and polars by using
 GINGA/LAC observations. He fitted the X-ray
 spectra by  single temperature bremsstrahlung model. Then WD masses were
 estimated by equaling  the resulting best fit bresstrahlung temperatures
 to the  maximal shock temperatures. In reality a significant part of the
 post-shock region has smaller temperatures due to radiative cooling and,
 as   menthioned by the author, WD masses obtained in this way are  lower
 limits only.

   In order to estimate the masses of WDs more carefully by 
 using X-ray spectra of IPs one needs to calculate the temperature and
 emissivity distribution in the post-shock region (PSR). The structure of
 the PSR was investigated  by Aizu (\cite{Ai}), Wu et al. (\cite{Wu}),
 Woelk \& Beuermann (\cite{Wo}), Cropper et al. (\cite{Cr99}), see also
 Frank, King \& Raine (\cite{FKR}). Wu et al. (\cite{Wu}) and Woelk \&
 Beuermann (\cite{Wo}) took into account  cyclotron cooling, which can be
 important for polars. In these papers it was shown that  cyclotron
 cooling is not important for IPs if the surface magnetic field is less
 than $\approx 10$ MG. The model of Wu et al. (\cite{Wu}) was used by
 Cropper et al. 
 (\cite{Cr98}, \cite{Cr99}) for fitting the spectra of IPs observed
 by GINGA/LAC and by Ramsay (\cite{R}) for fitting the spectra of
 IPs observed by RXTE/PCA. Beardmore et al. (\cite{BOH})  used the results
 of a simple analytical model of the PSR (see Frank, King \& Raine
 \cite{FKR}) for a comparison with  the spectra of V1223 Sgr observed by
 GINGA/LAC and ASCA. Ezucka and Ishida (\cite{EI}) estimated the WD
 masses of nine IPs using emission line ratios from ASCA observations.

The main
 systematic uncertainty in these works is caused by the fact that  neither
 the GINGA/LAC nor the RXTE/PCA detectors provide adequate information at
 energies $>$20-25 keV. Therefore,  it is hard to get solid estimates for
 the masses of the WDs if the mass is $>$ 0.6 M$_{\odot}$,  because in that
 case  the maximal post-shock temperature is $>$ 20 keV.

In this work we present results of WD mass estimates for 14 bright IPs,
obtained from broad-band (3-100 keV) RXTE spectra. Most of the sources
analized have a statistically significant signal at energies $>$10 keV. The
observed spectra have been fitted by PSR model spectra. We compare the WD
masses resulting from our  best fit parameters with the  WD masses that
were obtained by  other methods.

\section{The model}

The structure of the stationary post-shock
region in plane-parallel one-dimensional geometry is  described
(see e.g. Cropper et al. \cite{Cr99}) by the mass continuity equation

 \begin{equation}\label{mce}
\frac{d}{dz} (\rho v) = 0,
\end{equation}

the momentum equation

\begin{equation}\label{me}
\frac{d}{dz} (\rho v^2 + P) = -\frac{GM_{wd}}{z^2} \rho,
\end{equation}

the energy equation

\begin{equation}\label{ee}
  v\frac{dP}{dz} + \gamma P\frac{dv}{dz} =
     -(\gamma-1) \Lambda,
\end{equation}

and the ideal-gas law

\begin{equation}\label{igl}
P = \frac{\rho kT}{\mu m_H}.
\end{equation}
Here $z$ is spatial coordinate (see Fig.1), $v$ is the velocity of
matter, $\rho$ is the density, $T$ is the temperature, $P$  the gas
pressure,   $\gamma$=5/3 the adiabatic index,  and $\mu$=0.62 the mean
molecular weight of fully ionized matter of solar composition. The
cooling rate $\Lambda$ due to thermal optically thin radiation is:

\begin{equation}\label{cf}
\Lambda = \left ( \frac{\rho }{\mu m_H} \right )^2 \Lambda_N (T),
\end{equation}
where $\Lambda_N(T)$
is the cooling function, here taken for solar chemical composition as
calculated and  tabulated by Sutherland and Dopita (\cite{SD}).

Equation (\ref{mce}) has the integral
\begin{equation}\label{imce}
\rho v = a,
\end{equation}
where $a$ is the local mass accretion rate per unit area
(of dimension g cm$^{-2}$ s$^{-1}$). The local mass accretion rate is a
free parameter of the model. However, we found that the spectrum of
emergent radiation is rather insensitive to this parameter within
reasonable limits (i.e. as long as the emission region is optically thin
with respect to free-free scattering).

\begin{figure}
\includegraphics[width=\columnwidth,bb=0 166 355 509,clip]{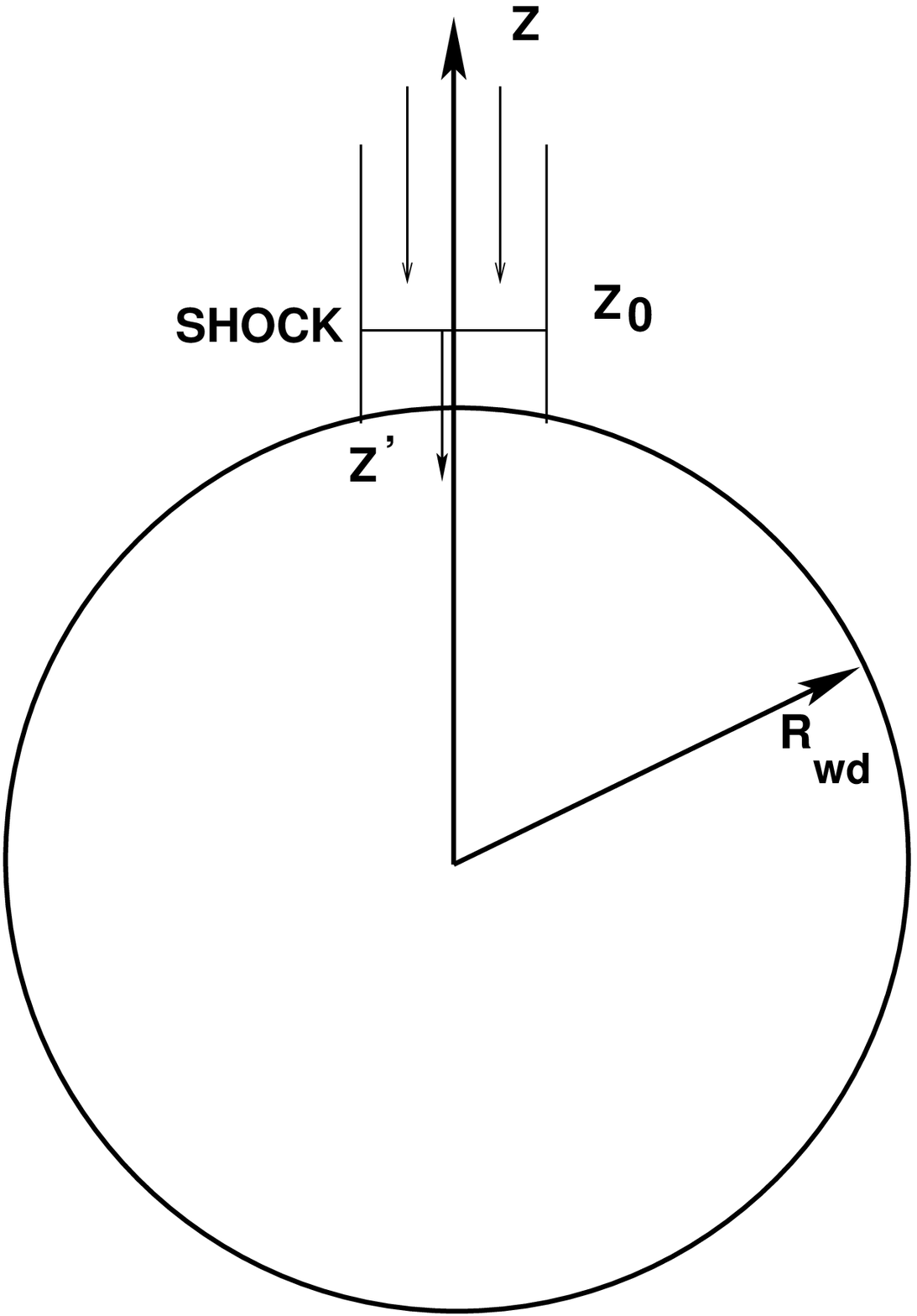}
\caption{\label{fig1}. Geometry of a
post-shock region model.}
\end{figure}

 Equations (\ref{me}) and (\ref{ee}) can be rewritten using
 (\ref{imce}) with the substitution $z'=z_0-z$, where $z_0$ is the shock
 coordinate (see Fig. \ref{fig1}):
\begin{equation}\label{dv}
 \frac{dv}{dz'}  = g(z') \frac{1}{v} - \frac{1}{a} \frac{dP}{dz'},
\end{equation}

\begin{equation}\label{dP}   \frac{dP}{dz'}  =
 \frac{(\gamma-1) \Lambda a + g(z') \gamma P   a/v}{\gamma P -  a v},
\end{equation}
 where
\begin{equation}{\nonumber}
g(z') =  \frac{GM_{wd}}{(z_0-z')^2}.
\end{equation}

Equations (\ref{dv}) and
 (\ref{dP}) were solved by the shooting method from $z=z_0$ ($z'=0$) to
 $z=R_{wd}$ ($z'=z_0-R_{wd}$) with boundary conditions at $z=z_0$:

\begin{equation}\label{v0}
   v_0  = 0.25 \sqrt{2GM_{wd}/z_0},
\end{equation}

\begin{equation}\label{rho0}
   \rho_0 = \frac {a}{v_0},
\end{equation}

\begin{equation}\label{p0}
   P_0 = 3 a v_0,
\end{equation}

\begin{equation}\label{t0}
T_0 = 3 \frac {\mu m_H}{k}  v_0^2.
\end{equation}

The WD radius was calculated from the Nauenberg
 (\cite{N}) WD mass-radius relation:
\[
    R_{wd} = 7.8 \cdot 10^8  \rm cm   \,\,\,\times
\]

\[
            \left [ \left (\frac {1.44
 M_{\odot}}{M_{wd}}\right )^{2/3} -   \left (\frac {M_{wd}}{1.44
 M_{\odot}} \right )^{2/3}  \right ]^{1/2}.
\]

In Fig.2 we present the
 temperature and the density profiles of the post-shock region resulting
 from our  model for a particular set of boundary values. For comparison we
 show the  profiles obtained by the simple analytical model (Frank, King \&
 Raine \cite{FKR}):
\begin{equation}\label{tz}
T(z) = T_0 \left ( \frac
 {z-R_{wd}}{z_0-R_{wd}} \right )^{2/5},
\end{equation}

\begin{equation}\label{rhoz}
\rho(z) = \rho_0 \left ( \frac
 {z-R_{wd}}{z_0-R_{wd}} \right )^{-2/5}.
\end{equation}
This model is
 based on the assumption of constant pressure in the post-shock region. In
 our model the pressure grows towards WD surface, therefore the temperature
 and the density in our model are larger.

   As  the RXTE/PCA  detectors
 have relatively poor spectral resolution and have sensitivity only at
energies higher than 3 keV we limited ourself to the  study of the
continuum emission.  The emergent lines at energies 6-7 keV  in the
resulting spectrum were mimicked by broad gaussian lines.

 The
emergent model spectrum is calculated by summing of the local
bremsstrahlung spectra:
\begin{equation}\label{isp}
F_E = \int^{z_0}_{R_{wd}} j(z) dz,
\end{equation}
where the local spectra were
taken in the following form (Zombeck \cite{Zo}):
\[
    j(z) = 9.52 \cdot 10^{-38} \,\,\,\times
\]
\[
\left( \frac{\rho(z)}{\mu
m_H}\right)^2   T(z)^{-1/2}\left( \frac{E}{kT(z)}\right)^{-0.4}
\exp\left(-\frac{E}{kT(z)}\right) .
\]

\begin{figure}
\includegraphics[width=\columnwidth]{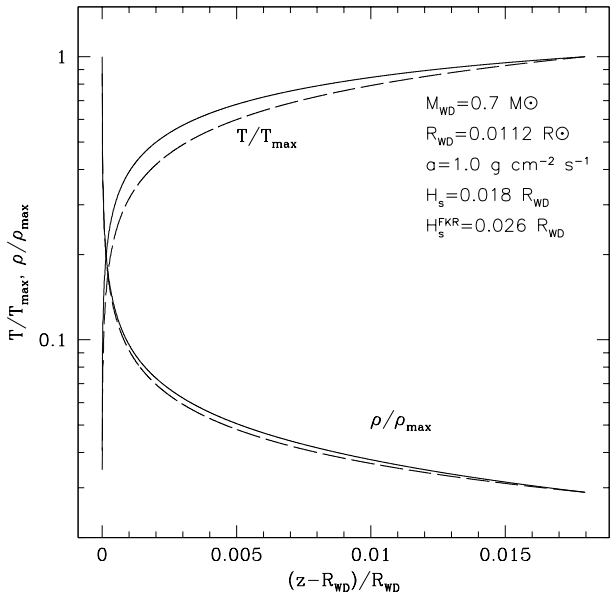}
\caption{\label{fig2}. Temperature and density profiles for one of the PSR
models. Dashed lines renote simplistic analytic solution of the structure
of post-shock region (Frank, King \& Raine \cite{FKR})}
\end{figure}

\section{Observations and data reductions}

 In order to compare the
results of our model calculations with observations we have chosen the data of
the RXTE observatory that cover a relatively  broad energy band  (3--250
keV), where IPs release most of their emission. We limited ourselves  only
to relatively bright IPs, as we are interested in the hard X-ray ($>$20
keV) part of the spectra of IPs.  The list of IPs consists of V1223 Sgr, FO
Aqr, EX Hya,  AO Psc, TV Col, GK Per, V709 Cas, PQ Gem, V2400 Oph, BG CMi,
V405 Aur,  V1062 Tau, DO Dra, TX Col.

 The data of the RXTE/PCA and the
RXTE/HEXTE were reduced by means of standard tasks of the LHEASOFT/FTOOLS
5.2 package. The background subtraction of the PCA detectors was done with
the CMl7\_240 model. The subtraction of the background of the HEXTE
detectors is especially important in our case which concentrates on the
hard X-ray part of the spectrum of IPs. In order to monitor the background
of the detectors the HEXTE is rocking by $\pm 1.5^\circ$ with respect to
the target.  We analised the HEXTE spectra of IPs only if the background
spectra obtained at the $+1.5^\circ$ and the $-1.5^\circ$ offset positions
were statistically compatible,  in other words if there were no
contaminating sources in the vicinity of our targets.

\section{Results}

\begin{figure*}[htb]
\hbox{
\includegraphics[width=0.9\columnwidth,bb=20 175 600 716,clip]{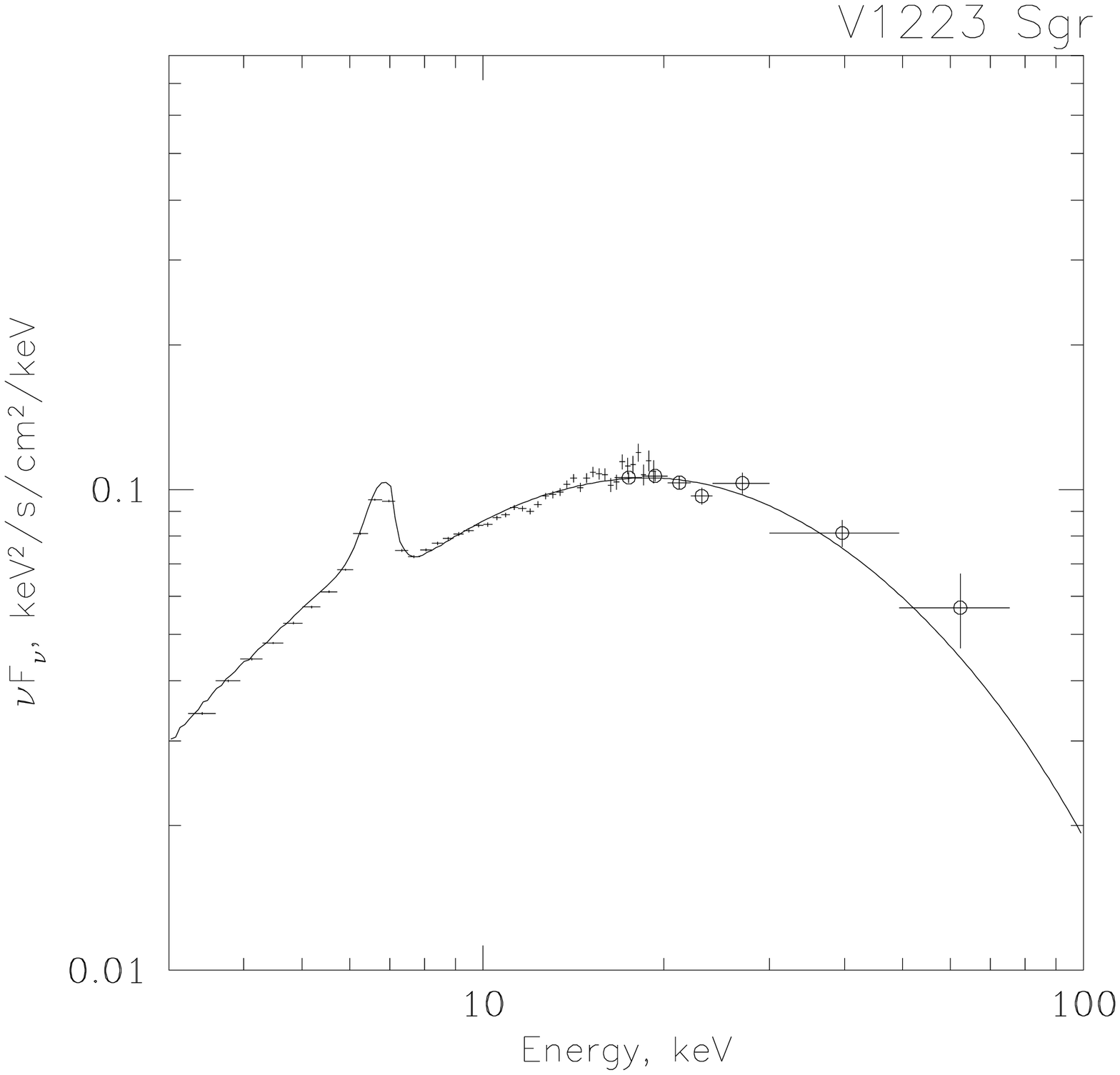}
\includegraphics[width=0.9\columnwidth,bb=20 175 600 716,clip]{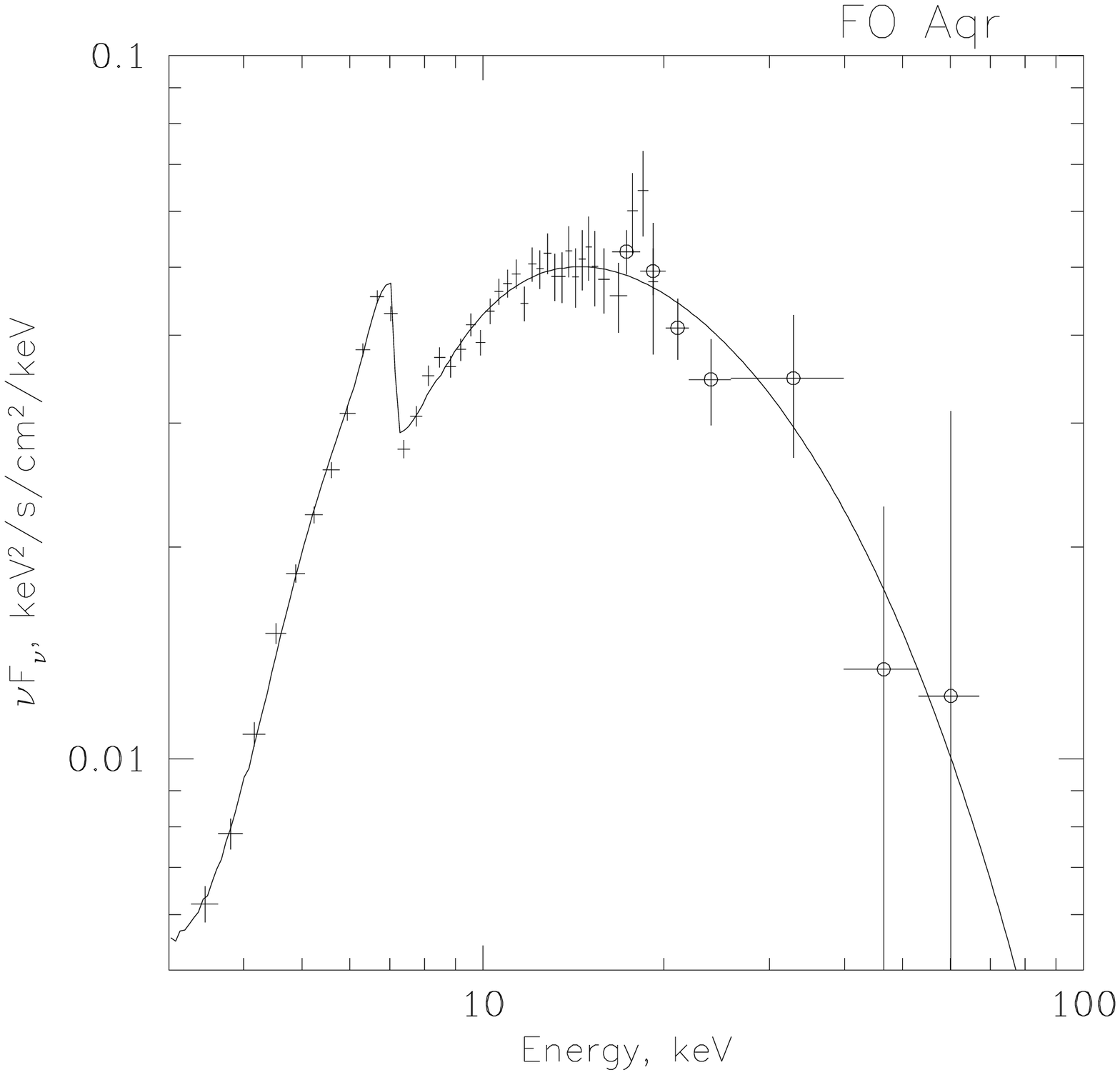}
}
\hbox{
\includegraphics[width=0.9\columnwidth,bb=20 175 600 716,clip]{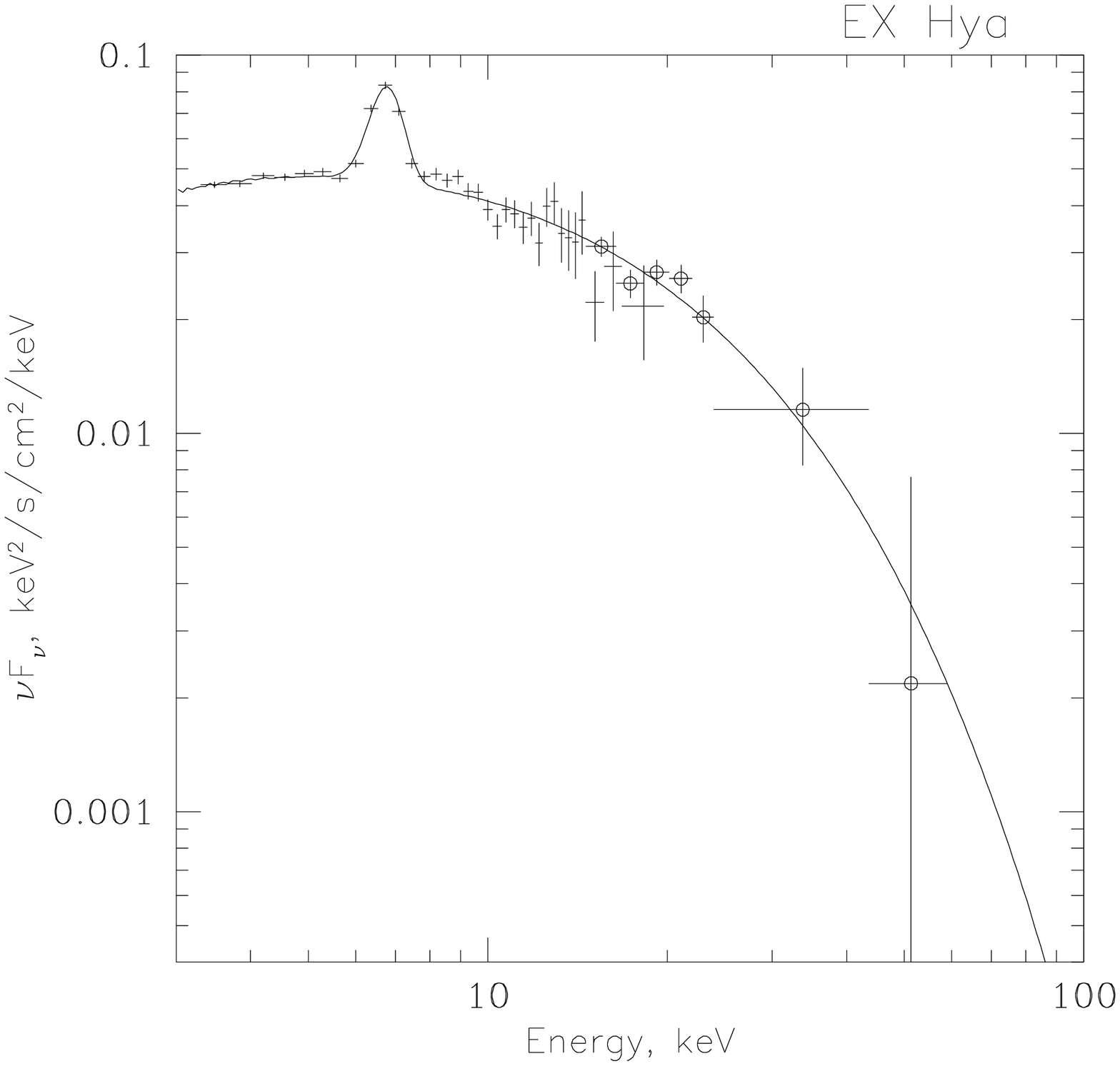}
\includegraphics[width=0.9\columnwidth,bb=20 175 600 716,clip]{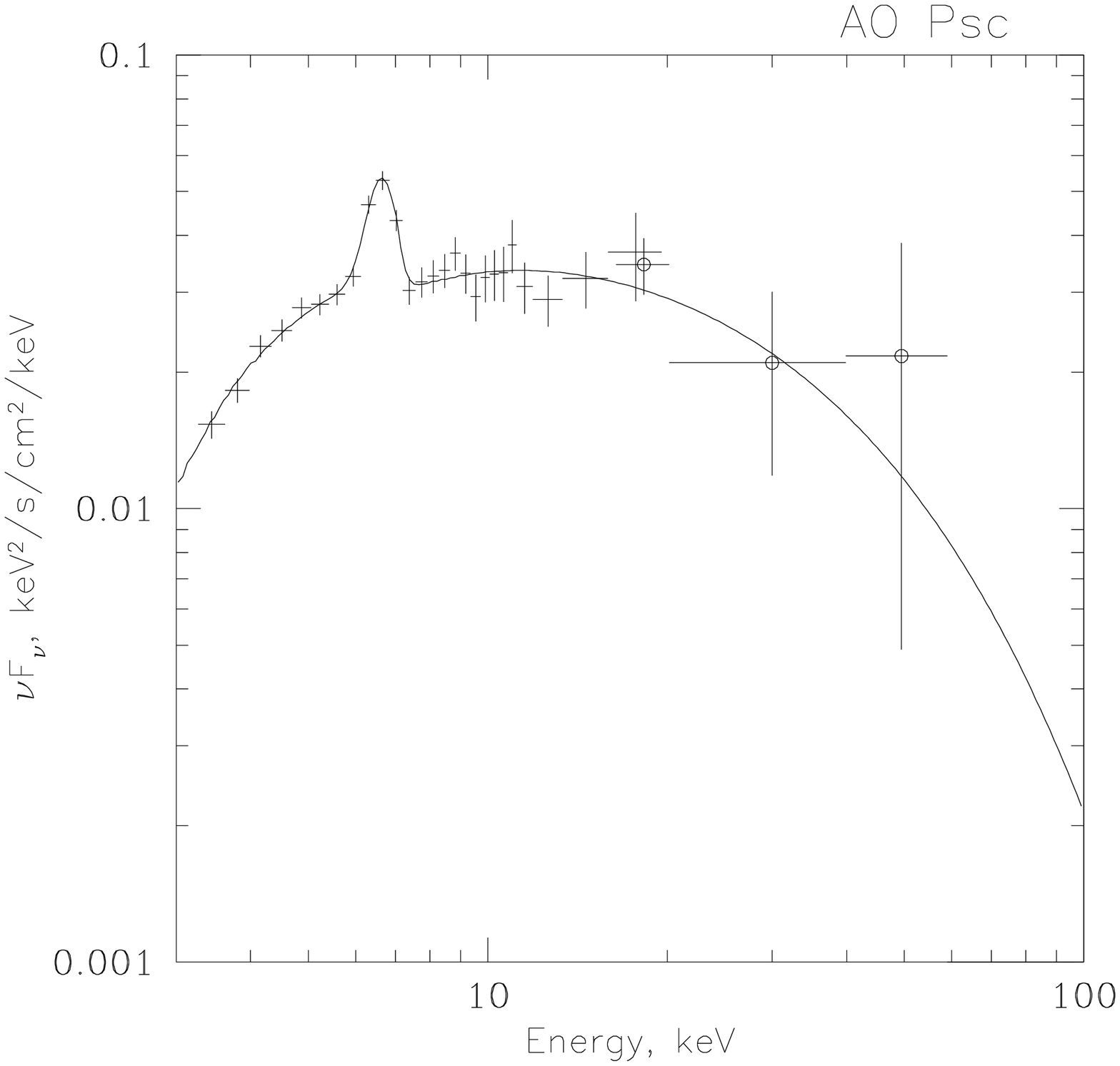}
}
\hbox{
\includegraphics[width=0.9 \columnwidth,bb=20 175 600 716,clip]{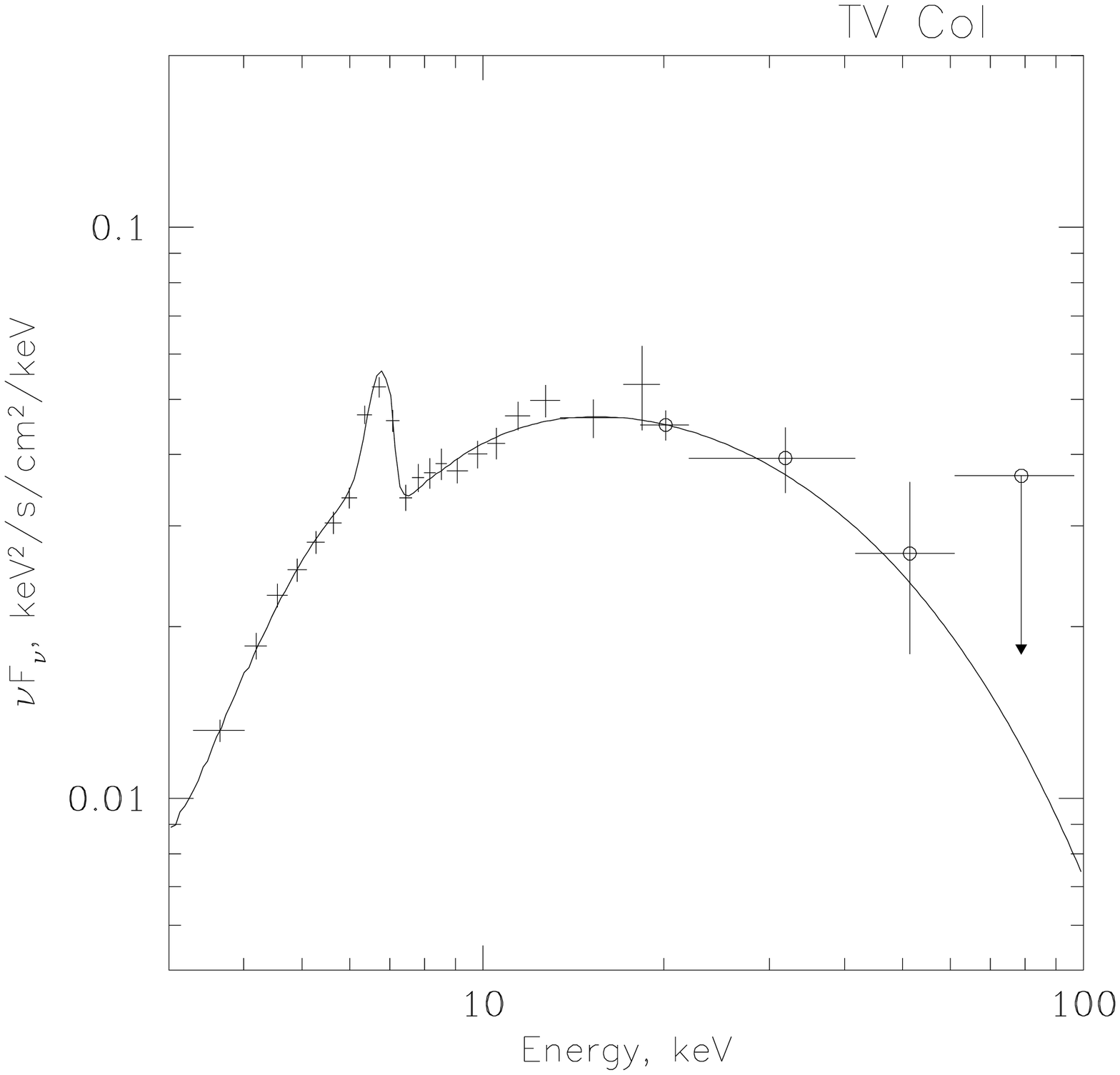}
\includegraphics[width=0.9\columnwidth,bb=20 175 600 716,clip]
{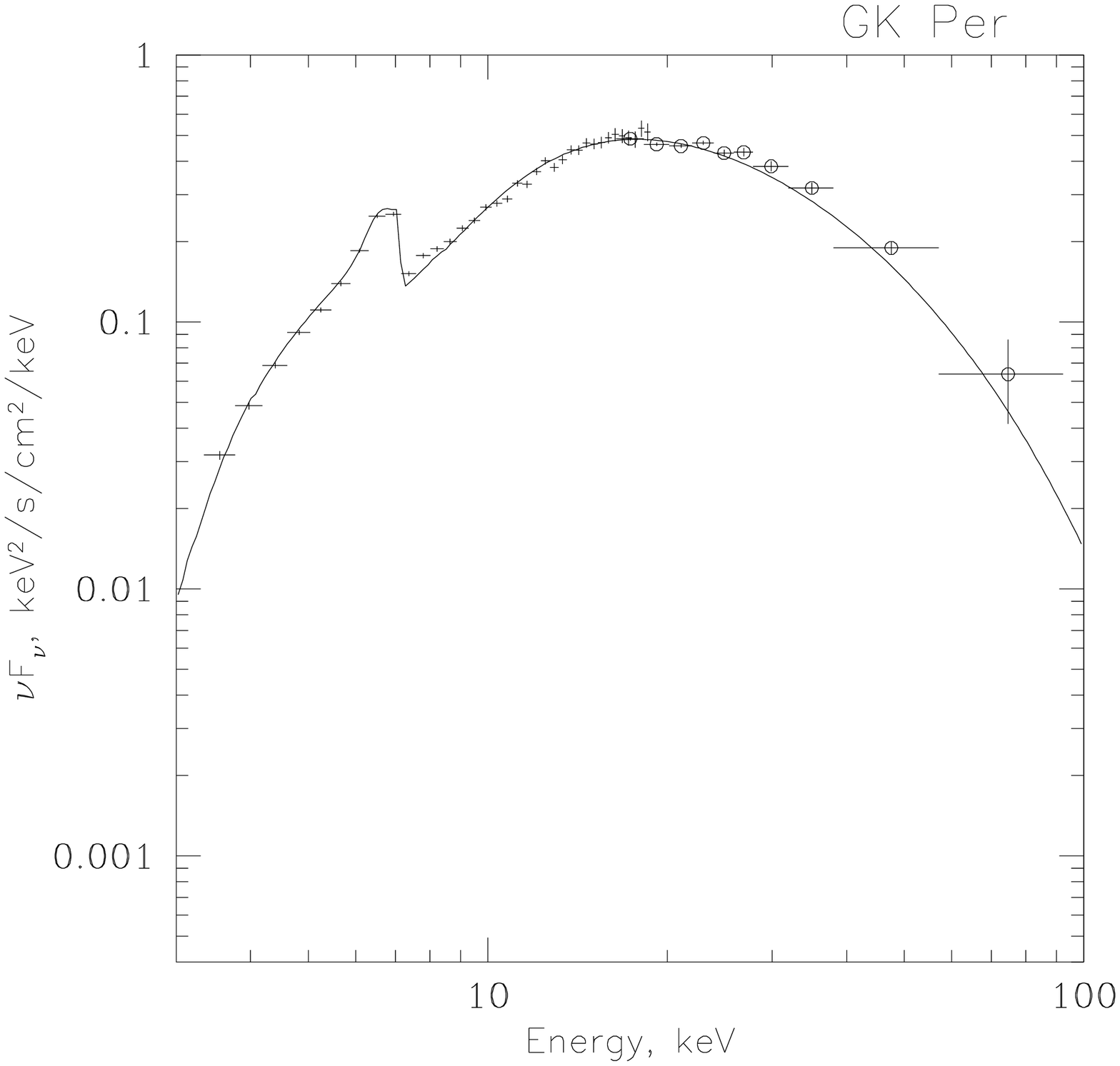}
}
\caption{\label{f3}Spectra of
intermediate polars obtained with the RXTE observatory. Crosses denote the
PCA data, open circles - HEXTE data.  Solid lines show best fit model with
parameters presented in Table 1}
\end{figure*}

The spectra of the 14 IPs, extracted from the RXTE data
are presented in Figs. \ref{f3} -- \ref{f4}.

 For fitting these spectra we
used results of our model calculations described  in the previous section
with one or two (only in the case of very bright IPs V1223 Sgr and GK Per) partial
covering components.  In contrast to the model used by Cropper et al.
(\cite{Cr98}) our model  lacks the reflected component. However, as has
been shown by Cropper et al. (\cite{Cr98}) taking into account the
reflection  changes the best fit parameters and thus the WD
mass only slightly.
In addition to that, the usage of spectral information at energies  higher
than $\sim$20 keV helps us to further reduce the influence of the
reflected component on the shape of the spectral cutoff.

  The best fit parameters resulting from our model are
presented in Table \ref{t1}. Using the observed and absorption-corrected
fluxes of IPs  we esimated their luminosities and mass accretion rates. The
best fit model is shown in Figs. \ref{f3} -- \ref{f4} by a solid line.

\begin{figure*}
\hbox{
\includegraphics[width=0.9\columnwidth,bb=20 175 600 716,clip]
{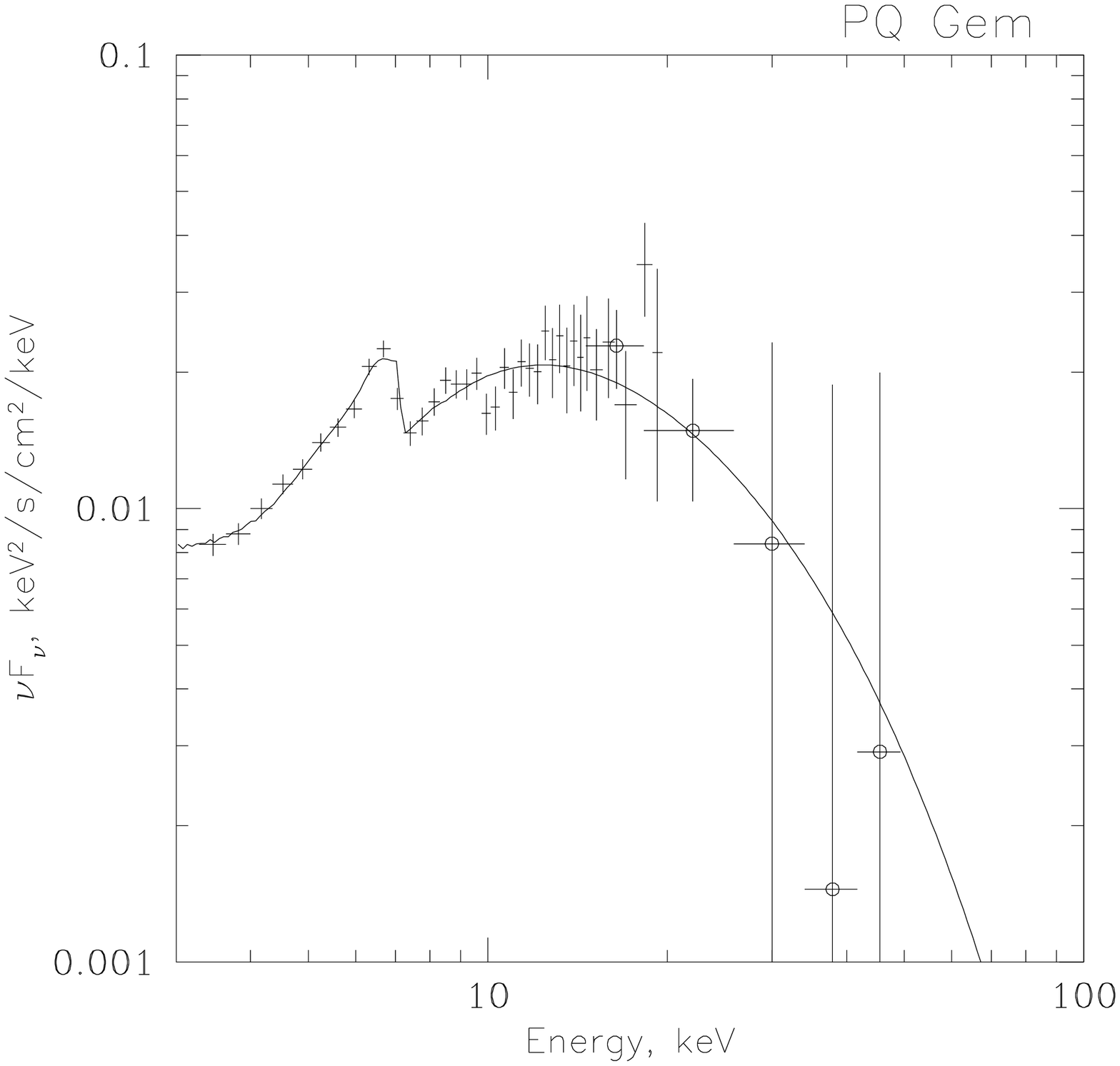}
\includegraphics[width=0.9\columnwidth,bb=20 175 600 716,clip]
{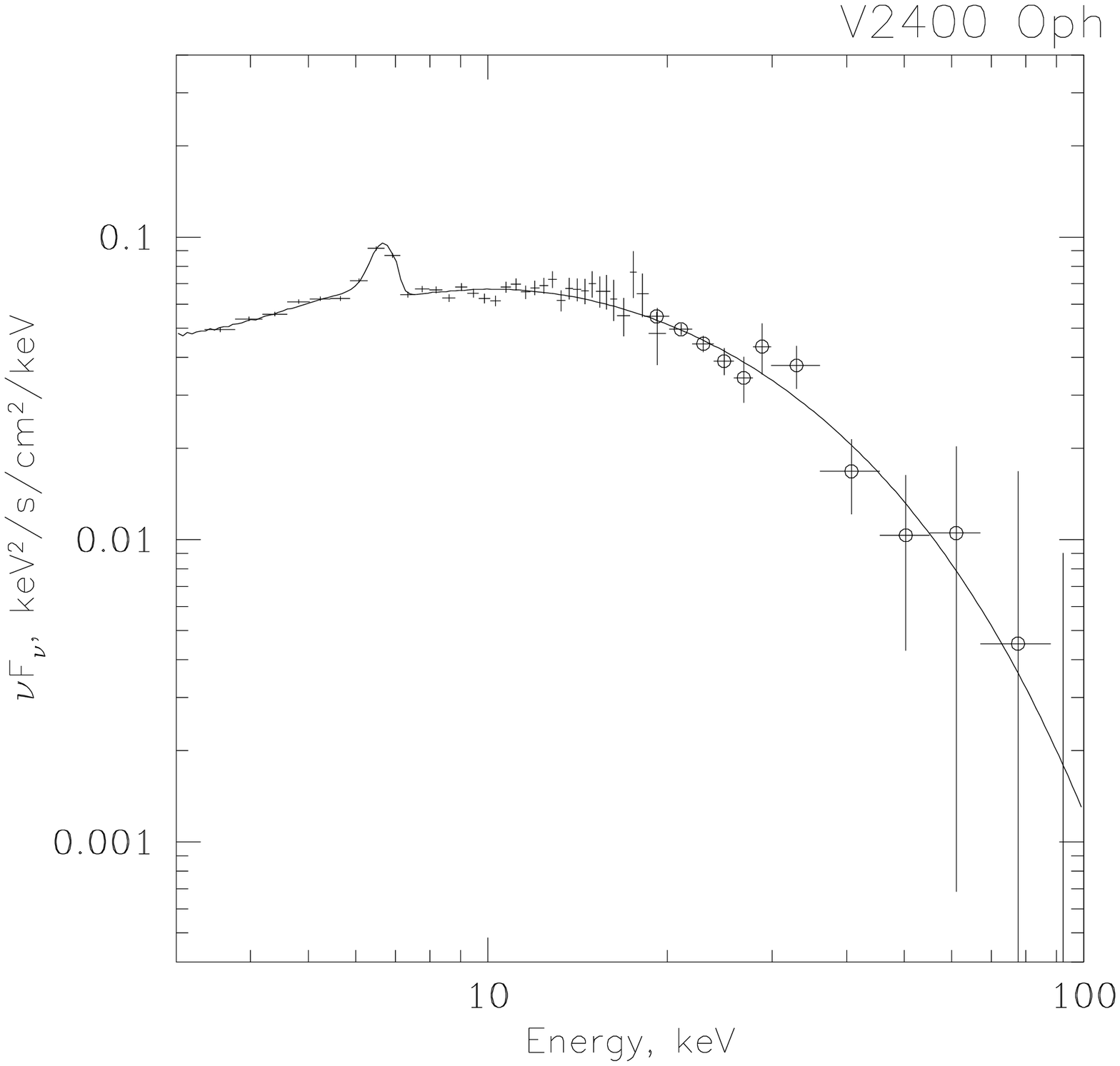}
}
\hbox{ 
\includegraphics[width=0.9\columnwidth,bb=20 175 600 716,clip]
{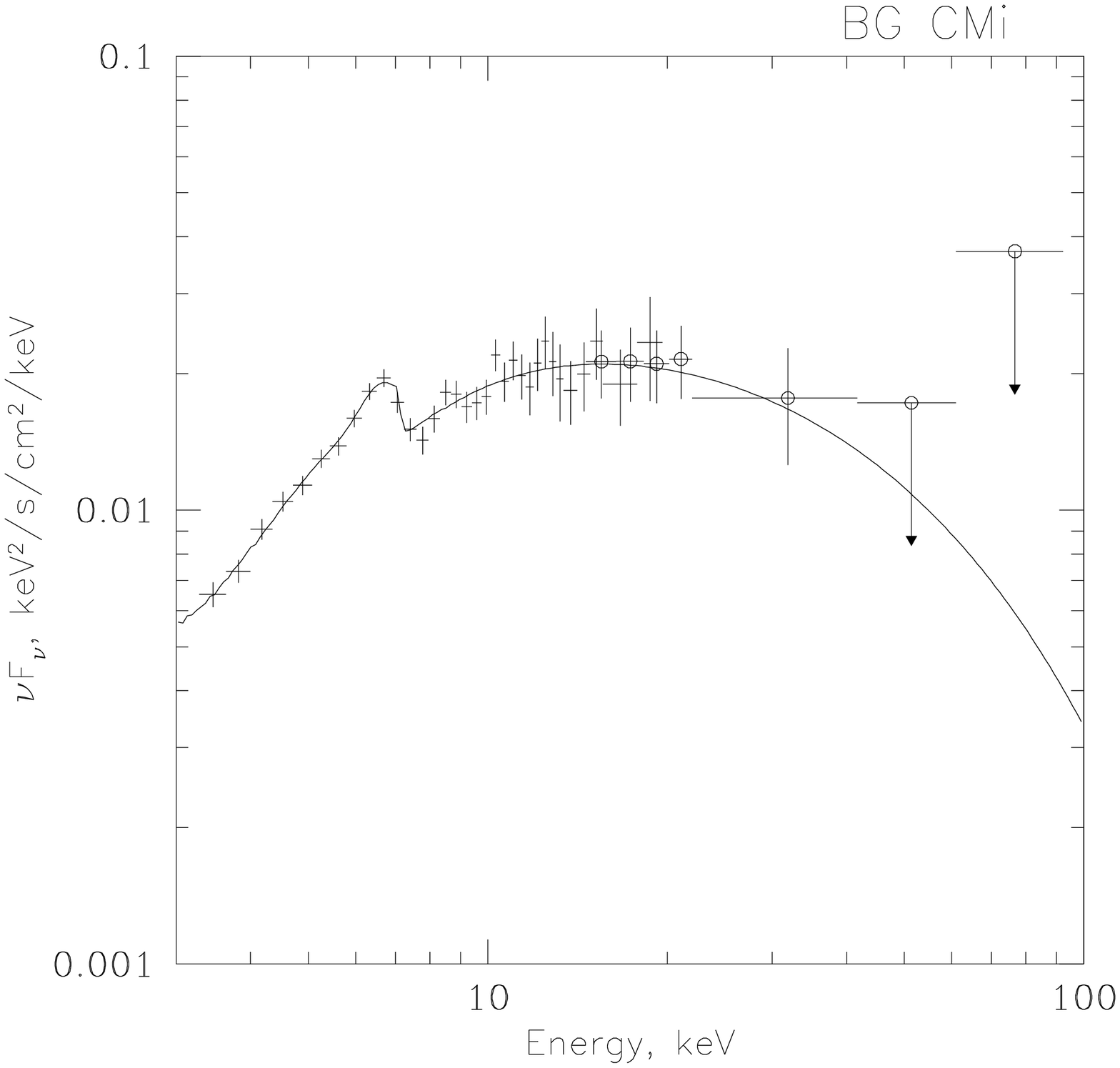}
\includegraphics[width=0.9\columnwidth,bb=20 175 600 716,clip]
{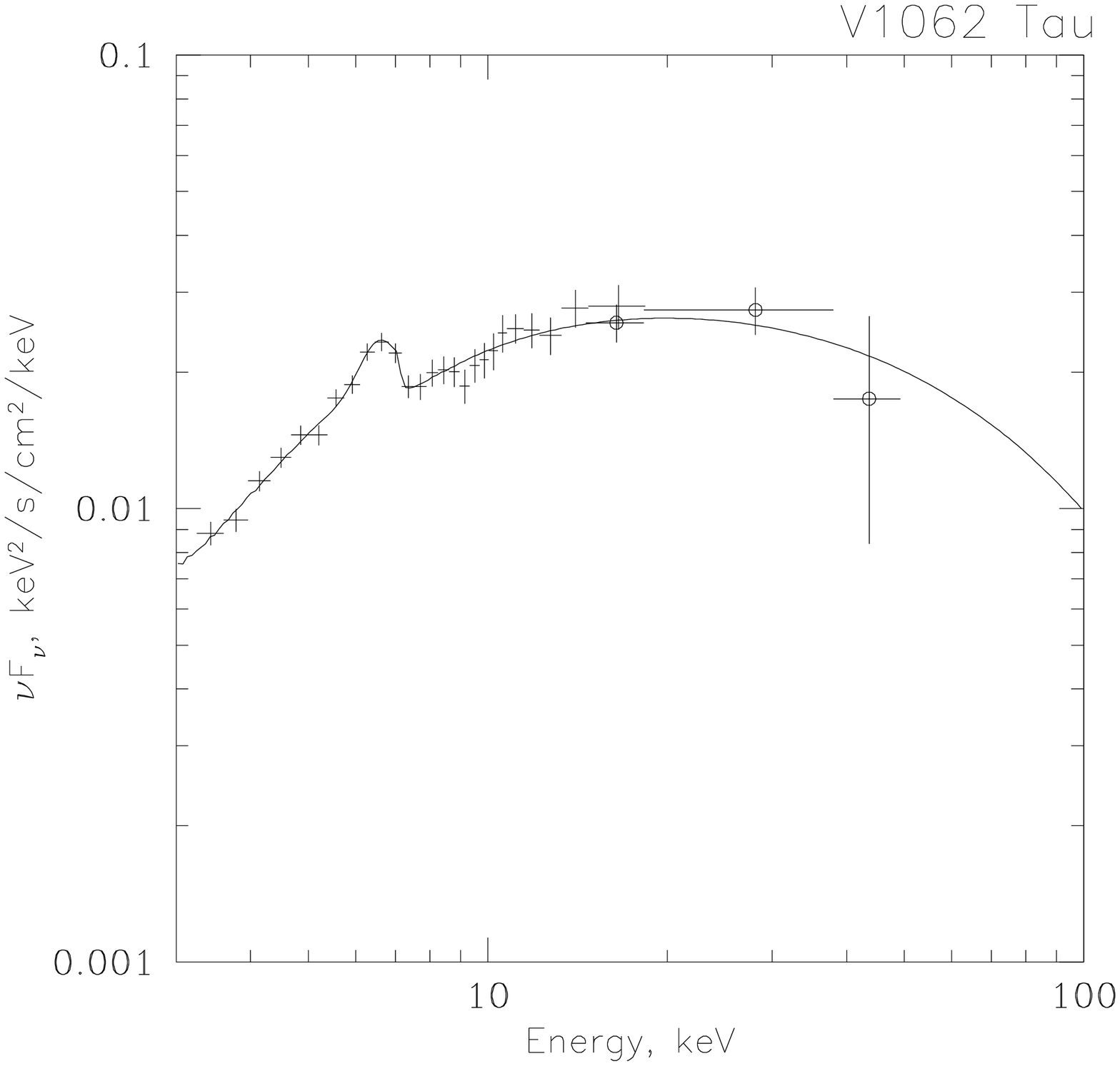}
}
\hbox{
\includegraphics[width=0.9\columnwidth,bb=20 175 600 716,clip]
{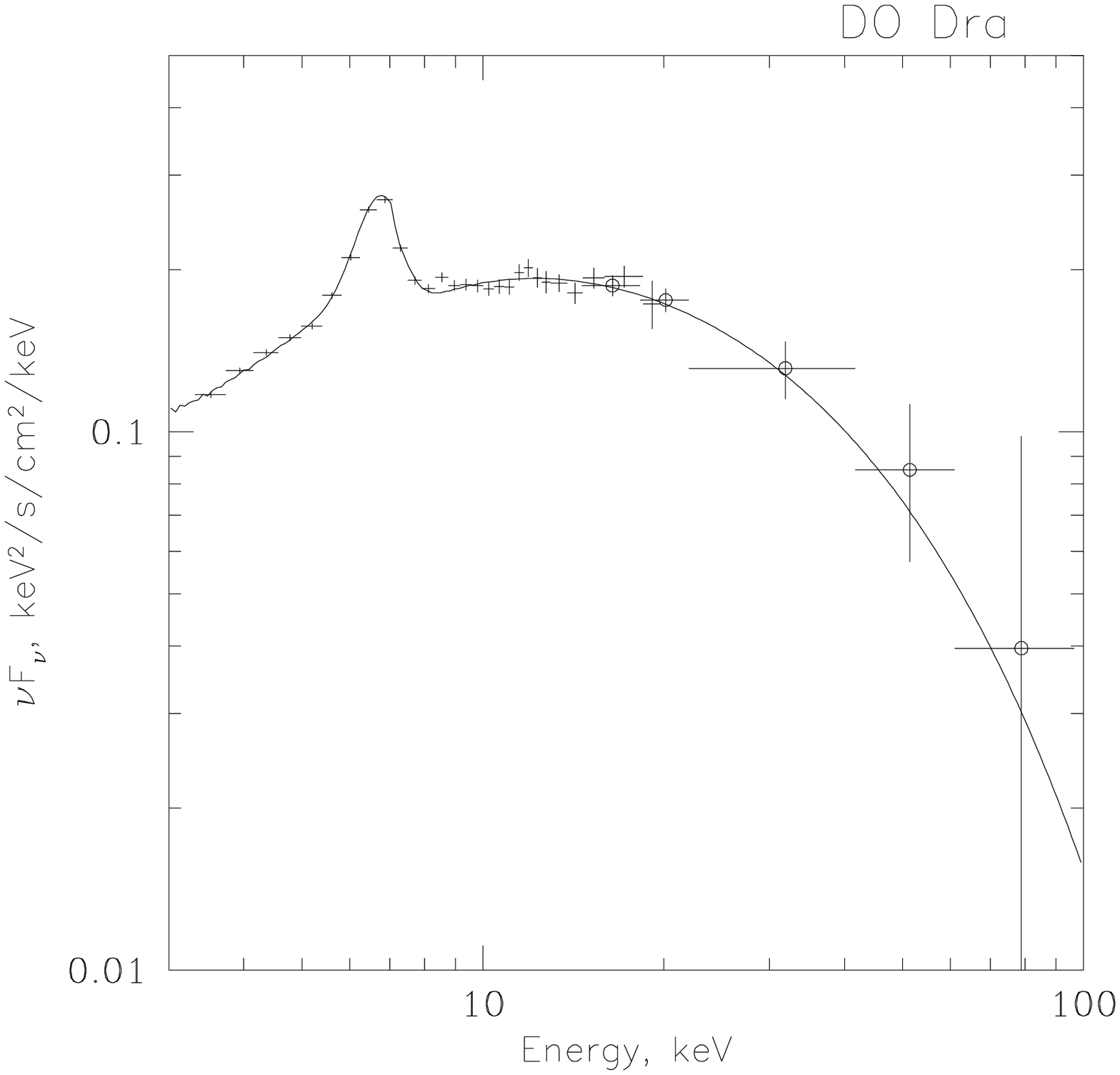}
\includegraphics[width=0.9\columnwidth,bb=20 175 600 716,clip]
{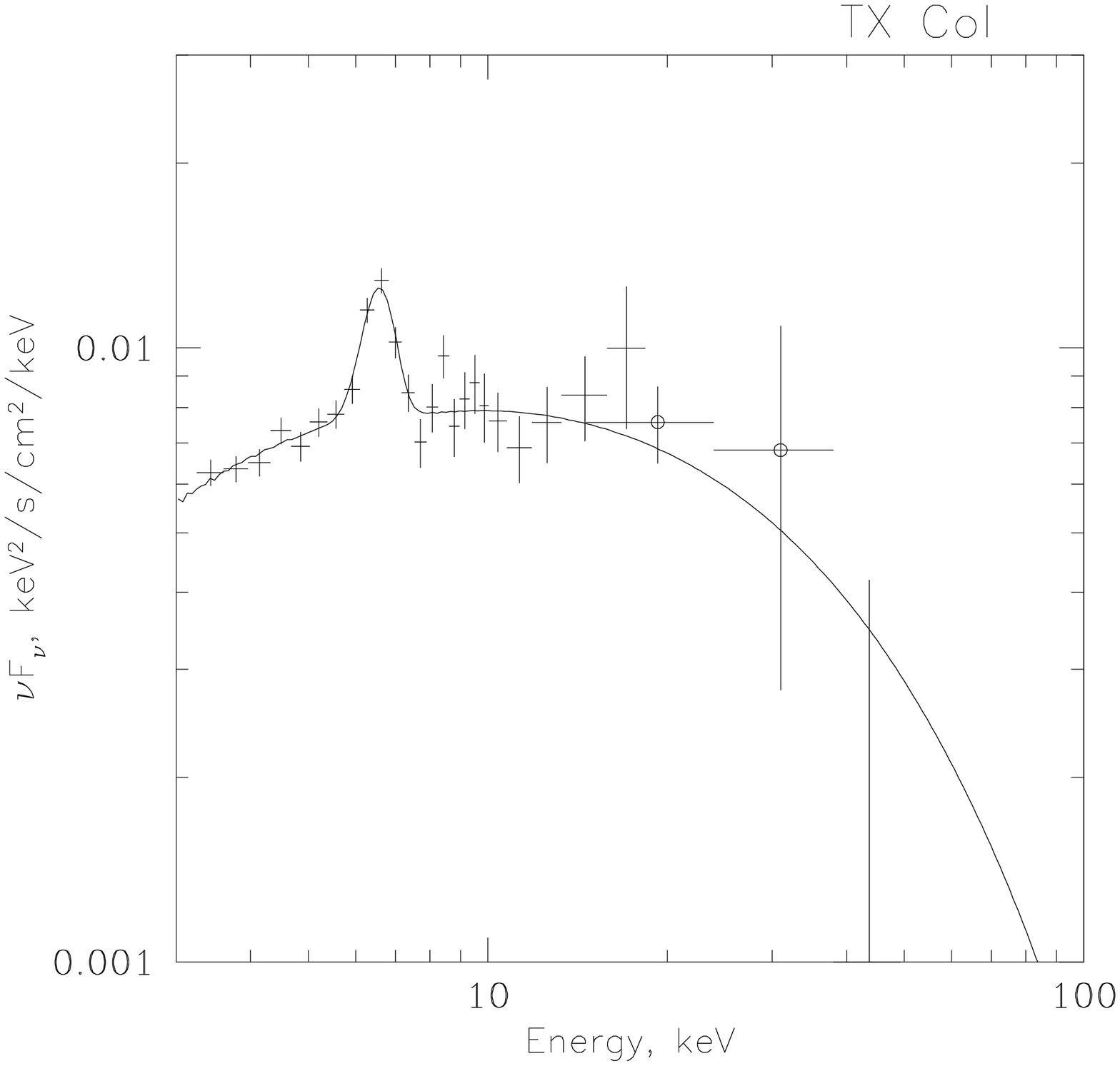}
}
\caption{\label{f4}The same as Fig. \ref{f3}}
\end{figure*}

\begin{figure*}
\hbox{
\includegraphics[width=0.9\columnwidth,bb=20 175 600 716,clip]
{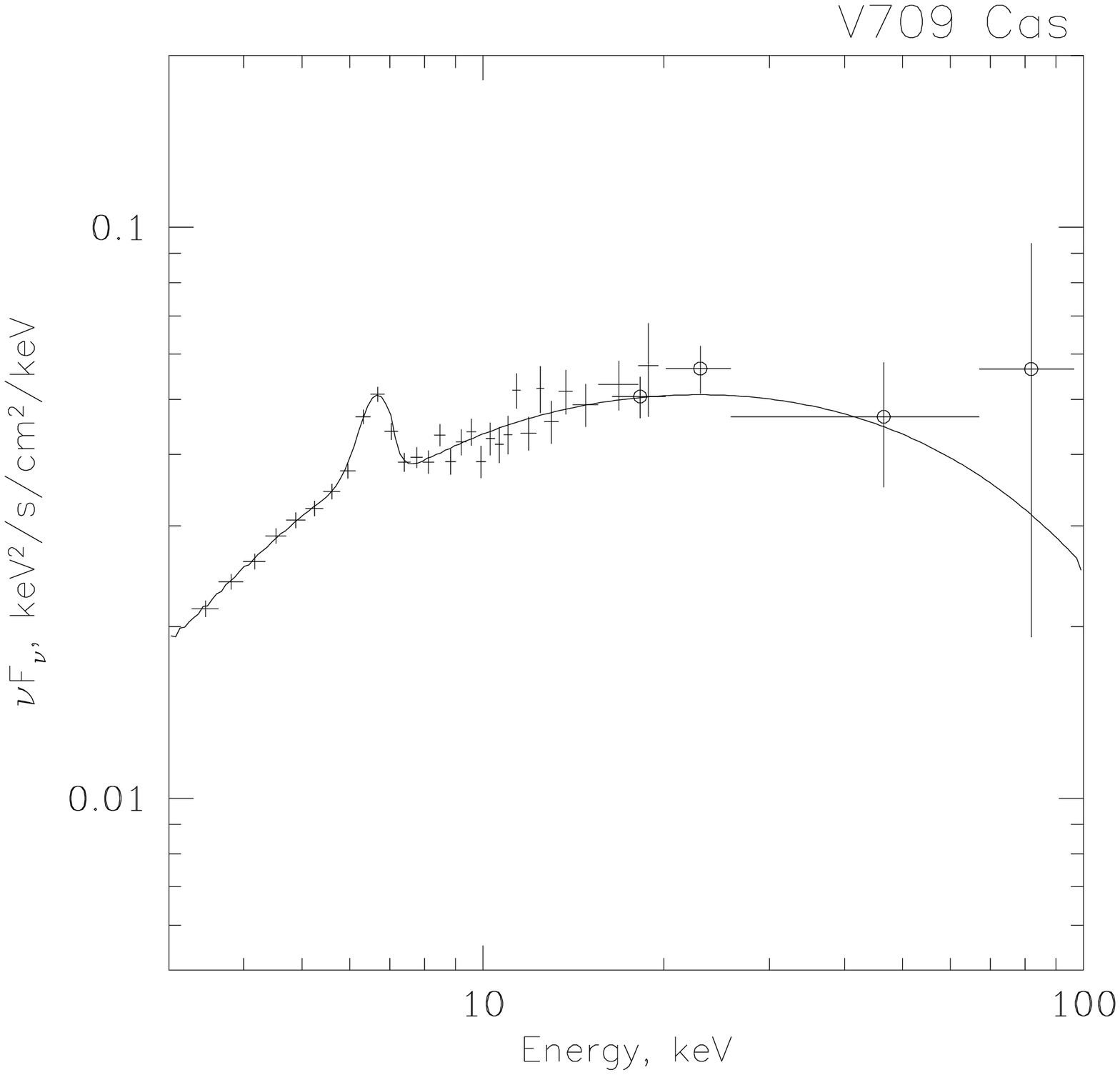}
\includegraphics[width=0.9\columnwidth,bb=20 175 600 716,clip]
{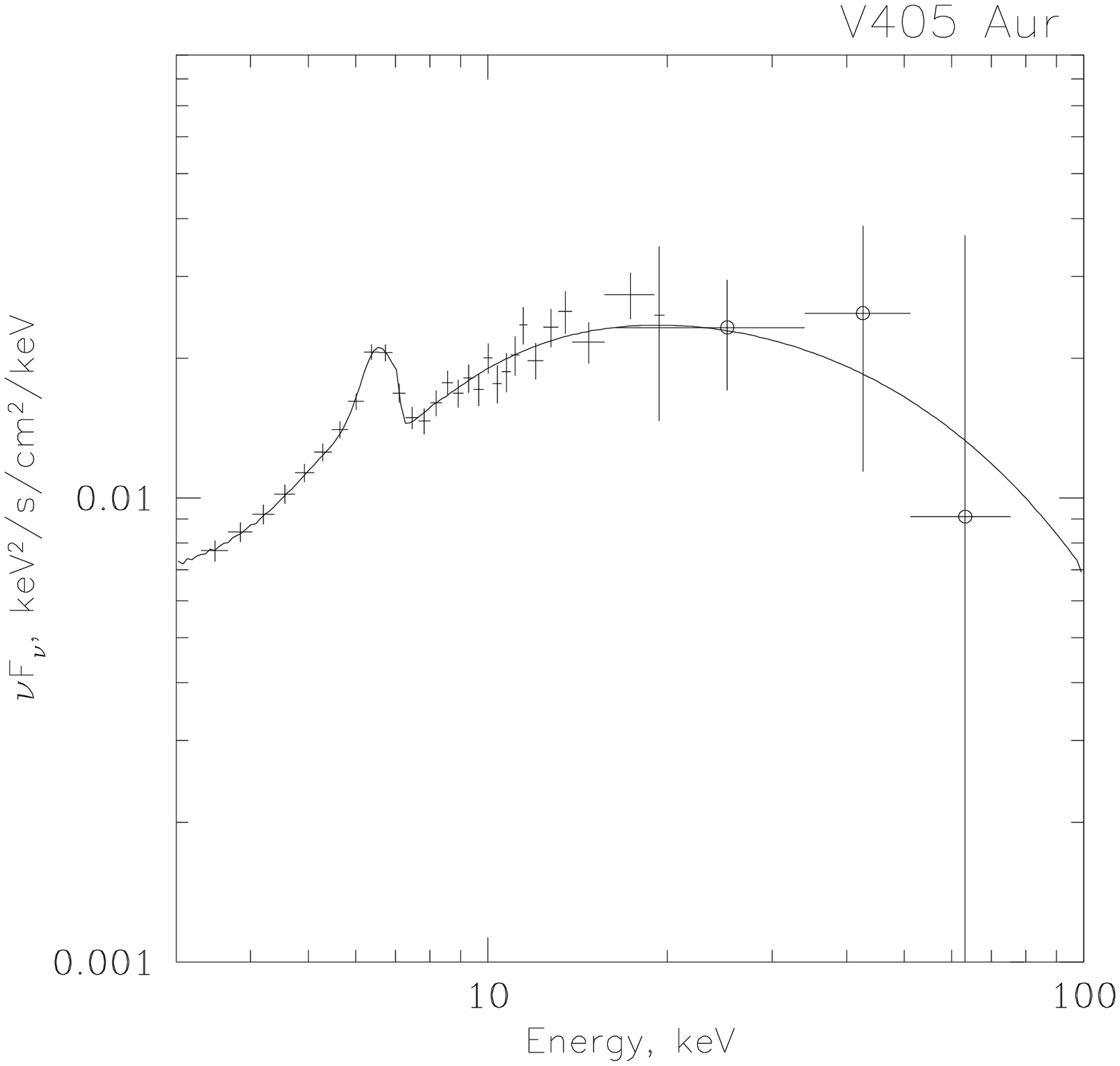}
}
\caption{\label{f4}The same as Fig. \ref{f3}}
\end{figure*}

\begin{table*}[htb]
\caption{\label{t1}Best fit model parameters  and inferred  physical system
parameters. Here C$_{\rm F}$ the partial covering coefficient, F$_{3-100}$
the observed X-ray flux in the band 3 -- 100 keV, F$_{0.1-100}$ the
unabsorbed X-ray flux in the band 0.1 -- 100 keV, both in 10$^{-11}$ erg
cm$^{-2}$ s$^{-1}$ units. Distances for V405 Aur, V709 Cas, PQ Gem and
V2400 Oph are assumed.}

\tabcolsep=3mm

\begin{tabular}{lllllllll}
\hline
Name & M$_{wd}$, & C$_{\rm F}$ & N$_H$, & F$_{3-100}$ & $F_{0.1-100}$ &
$\dot M$ & L$_x$ & d \\
    &  $M_\odot$  &       & $10^{22}$cm$^{-2}$&& & 10$^{16}$ g
s$^{-1}$ & 10$^{33}$ erg s$^{-1}$ & pc \\
\hline
V1223 Sgr & $0.95\pm0.05$ & 1.0$^a$ & $3.7\pm1$ &  4.2 &7.3 & 12.6 & 24.3 &
527$^1$\\
& & $0.57\pm0.05$ &  $28\pm3$&    &\\
FO Aqr & $0.6\pm 0.05$&$0.9\pm0.1$&
$36\pm4$    & 1.5&4.9 & 10.2 &9.4 & 400$^2$\\
EX Hya & $0.5 \pm 0.05$& 1.0
&  $0.6\pm1.0$  &1.5& 3.4 & 0.25 & 0.17 & 64.5$^3$ \\
AO Psc & $0.65\pm0.05$&$0.95\pm0.1$&  $10\pm5$      &  1.3 &2.5 &4.7 & 4.9
& 420$^4$  \\
TVCol & 0.84     & 0.81     &  21       &  1.7  &3.5 & 3.4 & 5.7 & 368$^5$ \\
GK Per   & $0.59\pm0.05$& 1.0 $^a$    &  $20\pm2$&    12.5& 88.6 & 81.5
&156 & 340$^6$\\
        &               &$0.75\pm0.05$ & $125\pm10$&\\
V709 Cas &  $0.9\pm0.1$ &$0.5\pm 0.1$&$29\pm5$&2.3&3.7& 5.76 & 11.1 &
(500)\\
PQ Gem   & $0.65\pm 0.2$ & 1.0      &  $5\pm1$    &  0.80&1.3 & 3.7 &3.9 &
(500)  \\
V2400 Oph& $0.59\pm0.05$ &$0.4\pm0.1$ &$29\pm10$&2.43&5.6 & 18.7 &16.75 &
(500)\\
BG CMi   & $0.85\pm0.12$ &$0.70\pm0.05$     &$25\pm6$&  0.77& 1.54 &5.3 &
9.0 & 700$^7$ \\
V405 Aur & $0.9\pm0.1$& $0.6\pm0.1$&$39\pm3$& 0.9 &1.7& 2.65 & 5.09 &
(500) \\
V1062 Tau & $1.0\pm0.2$& $0.6\pm0.1$ & $21\pm7$&  1.1&1.8 & 2.2 & 5.4&
500$^8$   \\
DO Dra   & $0.75 \pm 0.05$ & $0.4\pm 0.1$ &  $20\pm5$ &  7.8 &14.4 &3.1 &
4.1 & 155$^9$   \\
TX Col   & $0.7\pm0.3$ & $<0.4$ &  $<$60   &  0.32& 0.58 & 1.8 & 2.1 &
550$^{10}$   \\

\hline \\
\end{tabular}

$^a$ -- for systems V1223 Sgr and GK Per which have very high mass accretion
rates  we used two absorbers.

{\it References to distance estimates:} 1. Beuermann et al. (\cite{Beu}),
2. McHardy et al. \cite{McH} 3. Beuermann et al. \cite{Beu2},
4. Hellier, Cropper \& Mason \cite{HCM},5. McArtur et al. \cite{McA},
6. Warner \cite{W87}, 7. Berriman \cite{Be}, 8. Szkody \& Silber \cite{SS},
9. Mateo, Szkody \& Garnavich \cite{MSG}, 10.  Buckley \& Tuohy \cite{BT}
\end{table*}

\begin{table}[htb]
\caption{\label{t2}Comparision of mass estimates obtained in this work with
those obtained with the usage of other X-ray instruments}
\tabcolsep=1.5mm
\begin{tabular}{lllll}
\hline
Name   &  PCA+HEXTE  & RXTE/PCA & Ginga &  ASCA   \\
      &  $M_\odot$  & $M_\odot$ &$M_\odot$ & $M_\odot$  \\
\hline
V1223 Sgr    & 0.9  & 1.1      &        &  1.28  \\
FO Aqr       & 0.6  & 0.88     & 0.92   &  1.05   \\
EX Hya       & 0.5  & 0.45     & 0.46   &  0.48   \\
AO Psc       & 0.65 & 0.60     & 0.56   &  0.40   \\
TV Col       & 0.84 & 0.96     & 1.30   &  0.51   \\
GK Per       & 0.59 &          &        &  0.52   \\
V709 Cas     & 0.9  & 1.08     &        &         \\
PQ Gem       & 0.65 &          &        &         \\
V2400 Oph    & 0.59 & 0.71     &        &  0.68   \\
BG CMi       & 0.85 & 1.20     & 1.09   &         \\
V405 Aur     & 0.9  & 1.10     &        & $>$0.54 \\
V1062 Tau    & 1.0  & 0.90     &        &         \\
DO Dra       & 0.75 &          &        &         \\
TX Col       & 0.7  & 0.73     & 0.48   &  0.66   \\
\hline \\
\end{tabular}
\end{table}

\begin{figure}[htb]
\includegraphics[width=\columnwidth]{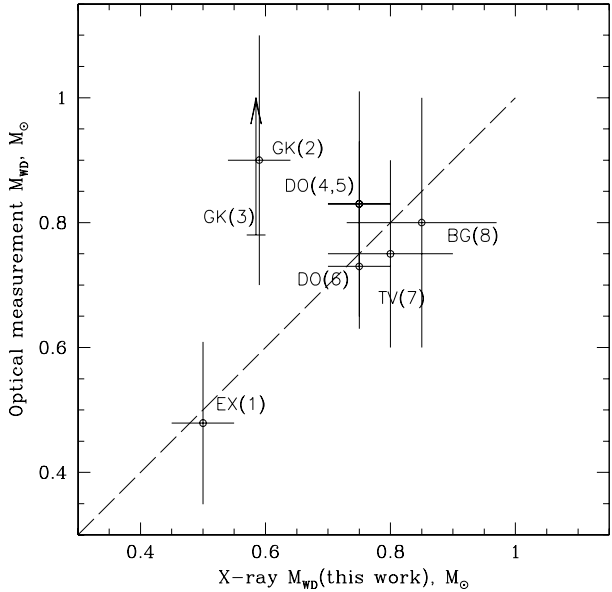}
\caption{\label{f5}Comparison of the masses of the white dwarfs as estimated
from  the RXTE  broadband spectra (this work) with estimates obtained
by other methods. (1)- Beuermann et al. (2003), (2) - Crampton et al. (1986),
 (3) -- Reinsch (1994), (4) -- Haswell et al. (1997),
(5) -- Mateo et al. (1991) , (6) -- Friend et al. (1990),
(7) -- Hellier (1993) , (8) -- Penning (1985)}
\end{figure}

We compare the resulting estimates of WD masses  with those measured
by other methods. In the catalog of Cataclysmic Variables
(Ritter \& Kolb \cite{RK}) there are only 5 intermediate polars that have a
spectroscopically measured mass. On Fig. \ref{f5} we present a comparison of
our mass estimates with those given in the catalog.  It is seen that we
have a relatively good agreement, however the error bars of all these
estimates are quite large. The only system that likely has a higher mass than
 we obtained from our X-ray spectral modelling is GK Per (during outburst).
One possible explanation of this fact could be the WD's fast
rotation ($P_{\rm sp}$ = 351 s).  It is well known that the radius of
the  magnetosphere in systems with an accretion disc depends on the
mass accretion rate.  If the WD in GK Per is rotating close
to its equilibrium period in its quiescent state (i.e. if the radius of its
magnetosphere is close to the corotation radius) then it means that during
an outburst (increase of the mass accretion rate by a factor of 10--20, see
 Ishida et al.  \cite{Ish2}) the size of the magnetosphere should shrink by
 a factor of $\sim 2$.  Therefore the radius of the magnetosphere in GK Per
during an outburst can be as small as $2-3\times 10^9$ cm, and thus, only  a
factor of 5--8 larger than the radius of the WD. In this case the
free fall velocity gained by the accreting matter near the surface of the
WD will be somewhat smaller than in the case of falling from infinity (our
model assumption).  Therefore the mass of the WD could  be
underestimated  by as much as 10-20\%.

Note  that our mass estimates are in
better agreement with those  obtained from optical methods than those
obtained by Cropper et al. (\cite{Cr99}) from their analysis of the GINGA/LAC
X-ray spectra ($\sim$2-20 keV) (see Table \ref{t2}). The masses obtained by
Ramsay \cite{R} from his analysis of the RXTE/PCA ($\sim$2-20 keV) spectra
are closer to our estimates, but also slightly higher. A possible reason for
this difference could be  the lack of the hard  X-ray part in the observed
spectra.

   It is interesting to mention that the parameters in our
Table \ref{t1} show a  correlation of the mass accretion rate with the
strength of photoabsorption, that, in principle, is anticipated for
IPs because  systems with higher mass accretion rates have higher
$N_H$ values. A similar effect was observed in  the case of GK Per
(quiescent and outburst spectra)  by Ishida et al. (\cite{Ish2}).

\section{Conclusions}

Estimating the masses of the WDs in cataclysmic variables is a
relatively complicated problem. The standard method based on radial
velocity measurements often gives unreliable results because of our poor
knowledge of the inclination angle of the binary system.  Besides, 
spectral lines from accretion discs are very broad in comparison with
Doppler shifts caused by orbital motion and their emissivity
distributions can be non-axisymmetric (in the frame of reference of
the WD). 
Therefore, additional assumptions, such as a relation between the mass
of the secondary star and the orbital period (based on a mass-radius
relation for main sequence stars, see Warner \cite{Wr}) are often used.

Magnetic cataclysmic variables provide us with an additional way for 
estimating the masses of the WDs. The matter accreted onto the WD
surface is heated in a standing shock and the maximal temperature of
the plasma after the shock front depends practically on WD mass only.
Therefore it is possible to estimate WD masses in polars and intermediate
polars by fitting their X-ray spectra with model spectra of the
radiation emerging from the post-shock region. If both methods of mass
determinations are available it is very interesting to compare the
resulting mass estimates. The temperatures of the PSR are relatively
high (10 -- 30 keV). Therefore the study of WD spectra in standard
X-ray band suffers from a lack of information at energies of a few 
$\times kT$, that would strongly help to constrain the WD mass 
estimate.

In this work we present the WD masses of 14 IPs estimated from fitting
the broad band (3-100 keV) RXTE/PCA+HEXTE X-ray spectra with model
spectra of the PSR. We did not take into account cyclotron cooling of
the PSR and therefore did not consider polars, where it can be important.
Our estimates of WD masses are smaller than those obtained from GINGA/LAC 
(Cropper et al. \cite{Cr99}) and RXTE/PCA (Ramsay \cite{R}) data. We also
find satisfactory agreement between our WD mass estimates and those
obtained by optical spectroscopic methods. We conclude that it is
necessary to include the hard X-ray spectral band (E $>$ 20 keV) for
obtaining reliable WD mass estimates in mCVs.

\begin{acknowledgements}  
The work was supported by the Russian Foundation of
Fundamental Research (grant $02$--$02$--$17174$), by the President
programm for support of leading science school (grant NSh - 1789.2003.2),
by grants of
Minpromnauka NSH-2083.2003.2 and 40.022.1.1.1103 and program of Russian
Academy of Sciences ``Non-stationary phenomena in astronomy''.
Research has made use of data obtained from 
High Energy Astrophysics Science Archive Research Center 
Online Service, provided by the NASA/Goddard Space Flight Center.

\end{acknowledgements}

\end{document}